\documentstyle[12pt,a4]{article}

\newcommand{\be}{\begin{equation}}
\newcommand{\ee}{\end{equation}}

\title{Classical solutions of the Gravitating Abelian Higgs Model.}
\author{Y. Brihaye, M. Lubo\\
Facult\'e des Sciences, Universit\'e de Mons-Hainaut,\\
B-7000 MONS, Belgium.}
\date{\ }
\begin{document}
\begin{titlepage}
\maketitle
\thispagestyle{empty}
\begin{abstract}
We consider the classical equations of the gravitating Abelian-Higgs model
in an axially symmetric ansatz. More properties of the 
solutions of these equations (the Melvin and the sting branches)
 are presented. These solutions are  also constructed 
 for winding numbers $N=2$.
 It is shown that these vortices exist in  attractive and repulsive phases,
separated by the value of the Higgs coupling constant parameter leading
to self-dual equations.
\end{abstract}
\vfill
\end{titlepage}
\newpage
\section{Introduction}
Most of the well known classical solutions in field theory are constructed in Minkowski space. Taking into account gravity, one recovers them as limiting cases but some new solutions appear which cannot be reached continuously from the existing one in flat space.
Such a situation occurs when one considers an Abelian-Higgs system. Imposing cylindrical
symmetry,one finds that there is only one solution~: the cosmic string \cite{niel}.
When gravity sets in, this configuration experiences a deformation which manifests itself as a conical singularity. As the angular deficit vanishes with the coupling constant measuring the gravitationnal interaction, this solution evolves smoothly from flat to
curved space. It has been shown recently that there exists another solution exhibiting axial symmetry in the gravitating Abelian-Higgs system \cite{verbin}. The profiles of
axial functions parametrizing the metric and  the matter fields have been obtained for
a unit winding number.

In this work, we first address the question of the link between the masses (inertial
mass and Tolman mass) of the new configurations and the parameters of the model
(the ratio $\alpha$ of the vector mass by the scalar mass on one side, the product
$\gamma$ of the Newton constant by the square of the Higgs-field's expectation value
on the other side). They are contrasted with those of the corresponding gravitating string. We then turn to  configurations with higher winding numbers and analyze their stability.
The phenomenon first pointed out in \cite{rebbi} is present for the gravitating strings
and for the new (Melvin-like) solution.

In a certain domain of the parameter space, the solution becomes closed i.e. 
defined within a maximal radius of the axial variable. We also 
investigate the dependence of this maximal radius versus $\alpha$ and $\gamma$.

\section{The equations.}
We consider the gravitating Abelian Higgs model in four dimensions. It is
described by the action
\be
S = \int d^4x \sqrt{-g}{\cal L}
\ee
\be
{\cal L} = {1\over 2} D_{\mu} \phi D^{\mu} \phi^{\ast} - {1\over 4}
F_{\mu\nu}F^{\mu\nu} - {\lambda\over 4} (\phi^{\ast} \phi-v^2)^2 + {1\over{16
\pi G}} R
\ee
where $D_{\mu} = \nabla_{\mu} - i e A_{\mu}$ is the gauge
covariant derivative, $A_{\mu}$ is the gauge potential
of the U(1) gauge symmetry,
$F_{\mu\nu}$ is the
corresponding field strength and $\phi$ is a complex scalar field with vacuum
expectation value $v$. The usual Ricci scalar is denoted $R$. We use the same
notations as \cite{verbin}.
\par In the following we study the classical equations
of the above field theory within the cylindrically symmetric ansatz. The
metric and matter fields are parametrized in terms of four functions of the
cylindrical radial variable  $r$  :
\be
ds^2 = N^2(r) dt^2 - dr^2 - L^2(r) d\phi^2-N^2(r) dz^2
\ee
\be
\phi = vf(r) e^{i n\phi}
\ee
\be
A_{\mu} dx^{\mu} = {1\over e} (n-P(r)) d\phi
\ee
Here $N,L,P,f$ are the four radial functions of $r$, $n$ is an integer
indexing the vorticity of the Higgs field around the $x_3$ axis. In \cite{verbin}
it was shown that the above ansatz consistently leads to the following
system of equations for the functions $N,L,P,f$~:
\be
\label{eq1}
{(LNN')'\over{N^2L}} = \gamma \left({P'^2\over{2\alpha L^2}}-{1\over 4}(1-f^2)^2
\right)
\ee
\be
\label{eq2}
{(N^2L')'\over{N^2L}} = - \gamma \left( {P'^2\over{2\alpha L^2}} + {P^2f^2
\over{L^2}} + {1\over 4} (1-f^2)^2\right)
\ee
\be
\label{eq3}
{L\over{N^2}} ({N^2P'\over L})' = \alpha f^2P
\ee
\be
\label{eq4}
{(N^2Lf')'\over{N^2L}} = f(f^2-1) + f {p^2\over {L^2}}
\ee
where $\alpha = e^2/\lambda$ and $\gamma = 8\pi G v^2$ are the physical constants.
The dimensionless coordinate $x = \sqrt{\lambda v^2}$ has been used and the primes
mean derivatives with respect to $x$.
The problem posed by the differential equations  is then
further specified by the following set of boundary conditions
\begin{eqnarray}
N(0) &=& 1\qquad , \qquad N'(0) = 0\\
L(0) &=& 0 \qquad , \qquad L'(0) = 1\\
P(0) &=& n \qquad , \qquad \lim_{x\rightarrow \infty} P(x) = 0\\
f(0) &=& 0 \qquad , \qquad \lim_{x\rightarrow \infty}f(x) = 1
\end{eqnarray}
which are necessary to guarantee the regularity of the configuration at the 
origin and the finiteness of the inertial mass  defined next.

\par The different solutions of the system can be characterized by the inertial
mass per unit length
\be
{\cal M}_{in} = \int \sqrt{-g_3} \ T^0_0 \ d x_1dx_2
\ee
and by the Tolman mass per unit length
\be
{\cal M}_{to} = \int \sqrt{-g_4}\  T^{\mu}_{\mu}\  dx_1 dx_2
\ee
In the cylindrically symmetric ansatz used and with the dimensionless
variable $x= \sqrt{\lambda v}r$ the two quantities above are given by the
following radial integrals
\begin{eqnarray}
G{\cal M}_{in} &\equiv& {\gamma\over 8} M_{in}\nonumber\\
&=& {\gamma\over 8} \int^{\infty}_0 dx NL \left( (f')^2 + {(P')^2\over{
\alpha L^2}} + {P^2f^2\over{L^2}} + {1\over 2} (1-f^2)^2\right)
\end{eqnarray}
\begin{eqnarray}
G{\cal M}_{to} &\equiv& {\gamma\over 2} M_{to}\nonumber\\
&=& {\gamma\over 2} \int^{\infty}_0 dx N^2L\left( {(P')^2\over{
2\alpha L^2}} - {1\over 4} (1-f^2)^2\right)
\end{eqnarray}
\par The integral determining the Tolman mass can be evaluated by means
of equation (\ref{eq1}) and is given by the following limit
\be
\label{toas}
G {\cal M}_{to} = {1\over 2}\lim_{x\rightarrow \infty} (LNN')
\ee
\section{Discussion of the global solutions.}
\par Let us first discuss the solutions of the system (\ref{eq1}), (\ref{eq4})
which are regular on $[0,\infty]$, i.e. when $N,L$ have no zero on this interval.
\subsection{The string branch.}
In the flat case $\gamma = 0$, equations (\ref{eq1})-(\ref{eq4}) admit the
celebrated Nielsen-Olesen \cite{niel} string as solution :
\be
\label{no}
N=1\quad , \quad L=x\quad , \quad P = P_{no}(x)\quad , \quad f=f_{no}(x)
\ee
where $P_{no}$ and $f_{no}$ are determined numerically. When the parameter
$\gamma$ is increased from zero, the solution (\ref{no}) gets progressively
deformed by gravity. In particular the functions $N,L$ become more complicated.
Asymptotically they obey the following behaviour
\be
\label{nas}
N(x\rightarrow \infty) = a
\ee
\be
\label{las}
L(x\rightarrow \infty) = bx+c\quad ,\quad  b > 0
\ee
here $a,b,c$ are constants depending on $\alpha, \gamma$. According to
(\ref{toas}), the Tolman mass vanishes.
\par For fixed $\alpha$, the set of solutions obtained by varying $\gamma$
(or vice versa) assemble into a branch of solutions called, after
\cite{verbin}, the string branch. They exist as global solution on $x\in
[0,\infty]$ up to a critical value $\gamma_{cr}(\alpha)$ (or, equivalently
if $\gamma$ is fixed up to $\alpha_{er}(\gamma))$ which is reached when the
parameter $b$ (defined in (\ref{las}) becomes zero (solutions corresponding
to $\gamma > \gamma_{cr}$ and $b<0$ will be the object of the next section).
 This phenomenon is illustrated by Fig.1 for $\alpha = 1.0$ (solid lines)
and $\alpha = 3.0$ (dashed lines).
\par The figure further indicates that the inertial mass $M_{in}$ increases
slightly with $\gamma$ while the parameter $b$ decreases linearly with $\gamma$.
Moreover, the critical value $\gamma_a(\alpha)$ increases with $\gamma$, e.g.
\begin{eqnarray}
\gamma_{cr}(1.0) &\approx& 1.66\\
\gamma_{cr}(2.0) &=& 2.0\\
\gamma_{cr}(3.0)&\approx& 2.2
\end{eqnarray}
\par The case $\alpha = 2$ possesses further algebraical properties because
there exist underlying self-dual (first order) equations implying the full
equations (\ref{eq1}), (\ref{eq4}). In particular we have in this case
\be
N(x) = 1.\qquad , \qquad M_{in} (\gamma) = 1
\ee
\be
a=1\qquad , \qquad b = 1-{\gamma\over 2}\qquad , \qquad \gamma_{cr}(2) = 2
\ee
\subsection{The Melvin branch.}
\par As discovered by \cite{verbin}, any global solution on the string
branch possesses a shadow solution. This second set of global solutions
of Eqs (\ref{eq1})-(\ref{eq4}) assemble into a branch named, after \cite{verbin},
the Melvin branch. The profile of a solution on the Melvin branch is
presented in Fig. 2 for typical values of the parameters $\alpha = 1.8, \gamma=1$.
\par In contrast to (\ref{nas}), (\ref{las}), the asymptotic behaviour of
the Melvin-branch solutions is such that
\begin{eqnarray}
N(x\rightarrow \infty) &\simeq& Ax^{2/3}\\
L(x \rightarrow \infty) &\simeq& B x^{-1/3}
\end{eqnarray}
As a consequence the Tolman mass is not zero any longer :
\be
G M_{to} = {1\over 2} \lim_{x\rightarrow \infty} (LNN') = {1\over 3} A^2B
\ee
\par The inertial and Tolman masses of the string and Melvin branches
are plotted on Fig. 3 as functions of $\alpha$ and for $\gamma=1$ (in this
case the critical value of $\alpha$ is $\alpha_{er} \approx 0.155$). The
figure clearly indicates that both (inertial and Tolman) masses are higher
for the Melvin branch than for the string branch. In the limit $\alpha
\rightarrow \alpha_{cr}$, the numerical analysis confirms that the inertial
massess (and also the Tolman masses) of the string and Melvin branches tend
to a common value (in the case of the Tolman mass, this value is zero).
\par The way the different radial functions associated with the two
solutions approach each other in the limit
$\alpha \rightarrow \alpha_{cr}$ is illustrated by Figs 4 and 5 where the profiles
of the functions are presented respectively for $\alpha = 0.2$ and $\alpha
= 0.16$. In particular, it is seen on the figures that the matter functions
$f,P$ of the string branch deviate rather slowly from their conterparts of the
Melvin branch.
\par The evolution of the masses $M_{in}, M_{to}$ for $\gamma\rightarrow 0$
on the Melvin branch is summarized by Fig. 6. The radial integral defining
these quantities clearly diverges when $\gamma$ approaches zero but the
Einstein-Maxwell equations can be recovered from Eqs.
(\ref{eq1})-(\ref{eq4})  by rescaling the radial variable $x$ and
the function $L(x)$ according to
\be
x = \sqrt{\gamma} y\quad , \quad L = \sqrt{\gamma\over{\alpha}} \tilde L
\ee
and setting $\gamma=0$ afterwards. Equation (\ref{eq4}) then 
decouples and the remaining equations are the
Einstein-Maxwell equations in the axially symmetric ansatz. 
The profile of the Melvin solution is
illustrated on Fig. 7. It has
\be
G {\cal M}_{in} = {1\over 4}
\ee
independently of $\alpha$.
\par This situation is somehow reminiscent to the case of the gravitating
sphaleron solution of the Einstein-Weinberg-Salam equation (EWS)
\cite{volkov}, \cite{bd}. The role played here by the flat Nielsen Olesen is
played in EWS equation by the flat klinkhamer-Manton Sphaleron \cite{km} while
the role of the Melvin solution on the top of the upper branch is played
by the first solution of the Bartnik-McKinnon series of solutions
\cite{bm}. However, unlike in the present theory, the transition from the
sphaleron to the Bartnick-Mc Kinnon solutions in the EWS model is smooth \cite{bd}.

\subsection{Solutions of vorticity $>1$.}
\par Coming back to the flat Nielsen-Olesen equations, we know that solutions
of vorticity $n$ can be constructed while imposing 
the following boundary conditions on $P(x)$ :
\be
P(0) = n\quad , \quad P(\infty) = 0
\ee
\par Denoting by $M(n)$ the classical energy of the Nielsen-Olesen solution
of vorticity $n$, one of the distinguished feature of the self dual case
($\alpha = 2$) is the mass relation
\be
M(n) = n M(1)
\ee
It is therefore a natural problem to check if the relation above still holds
in presence of gravity. We did so by chosing $\gamma=1$ and by studying
the inertial mass (which directly generalize the classical energy of the
flat case) for the values of $\alpha$ close to $\alpha=2$. These results
are presented on Fig. 8. The quantities $M_{in}(1)$ and ${1\over 2} M_{in}(2)$
cross at the self dual point $\alpha=2$ and their behaviour demonstrates
that the Rebbi-Jacobs phenomenon \cite{rebbi} also holds when gravity
is added. Namely for $\alpha > 2$ (resp. $\alpha <2$) the binding energy of
the $n=2$ string solution is negative (resp. positive).

\section{Closed solutions.}
As said above, the critical value $\gamma_{cr}(\alpha)$ (or equivalently
$\alpha_{cr}(\gamma)$) is determined when the slope parameter $b$ of the
function $L$ si zero. For $\gamma > \gamma_u(\alpha)$ (or $\alpha < \alpha_u(
\gamma))$ the string and Melvin branches of solution continue to exist but
they are not any longer available as regular solution for $x\in [0,\infty]$.
The fact that one of the functions $N$ or $L$ has a zero at, say $x=x_0$,
leads to a singular point of Eq. (\ref{eq1}) or (\ref{eq2}). Correspondingly
the solution can only be constructed for $x\in [0,x_0]$.
\par The inertial masses of the solutions continuing the Melvin and string
branches for $\alpha < \alpha_{cr}$ are presented in Fig. 3. Clearly, the
closed solution having the lowest mass is the continuation (for $\alpha
<\alpha_{cr}$) of the Melvin branch. In \cite{verbin}, it is called the
Kasner branch. An example of such a solution is presented on Fig. 10
for $\gamma=1, \alpha = 0.14$. It terminates at $x_0 \simeq 26$. The
evolution of $x_0$, the maximal value of $n$, in function of $\alpha$ is
presented on Fig. 11.

\section{Conclusion and outlook}
We have seen that, for a fixed $\alpha$, the mass of a string grows with $\gamma$
while the Melvin-like configuration display the opposite behaviour. This results in the existence of a critical value $\alpha_0$ such that the new solution is heavier 
(resp. lighter) than a string for $\alpha > \alpha_0$ (resp. $\alpha < \alpha_0$).
The numerical calculations show that $\alpha_0 = 0.5$. Another particular value of
interest is $\alpha = 2$, for which there exists first order equations implying the
 classical equations (this is self duality).
For $\alpha > 2$ the Melvin-like configurations have a negative binding energy
and, likely, are stable versus decaying into several solutions of unit vorticity.
This is also the case with cosmic string solutions. So the kind of solutions 
considered here obey the Jacobs-Rebbi \cite{rebbi} phenomenon
on both sides of the self
dual point $\alpha = 2$.
We also studied how the radius of the closed solutions grows with $\alpha$.

The Melvin-like solution raises many questions. A first one is a full 
understanding of its geometry \cite{frolov}. A second one is the elaboration
of a plausible mechanism for the generation of such a configuration. For example,
cosmic strings can be formed during a phase transition. This scenario seems
conceptually difficult to implement here since one would have to perform quantum
field theory at finite temperature and on a curved background. 

Cosmic string can also be formed by the collision of two monopoles. If the Melvin-type
configuration has to be generated in this way, then the problem is a technical one
since one has to solve the field equations for a fully time and position dependant
ansatz. If any of these two mechanisms takes place with a reasonnable efficiency, then
the new solution is likely to play a significant role in cosmology, like its string
cousins whose possible implications ranges from density generation to baryogenesis
\cite{jeannerot}.

The kind of calculations reported here could be extended to
other members of the Abelian models hierarchy\cite{tcha}
for which self dual equations are available as well.  
Lagrangians constructed by superposing a few members of the hierarchy
could also be considered \cite{arthur}; for such models, however, 
(and at least for flat space) only quasi self dual equations exist.

\newpage
\centerline{Figures Captions}
\begin{description}
\item [\ ] {\bf{Figure 1}}
The inertial mass $M_{in}$ and the quantities $a,b$ defined in (\ref{nas}),
(\ref{las}) as functions of $\gamma$ and for $\alpha = 1,0$ (solid)
and $\alpha = 3.0$ (dashed).
\item [\ ] {\bf{Figure 2}}
The profiles of the solution on the Melvin branch for $\alpha = 1.8, \gamma
= 1.0$.
\item [\ ] {\bf{Figure 3}}
The intertial mass $M_{in}$, the Tolman mass $H_{to}$ and the quantities
$a,b$ of Eqs. (\ref{nas}), (\ref{las}) as functions of $\alpha$ (for
$\gamma=1$) for the string (solid) and Melvin (dashed) solutions.
\item [\ ] {\bf{Figure 4}}
The string (solid) and Melvin (dashed) solutions for $\gamma=1.0$,
$\alpha = 0.2$.
\item [\ ] {\bf{Figure 5}}
The string (solid) and Melvin (dashed) solutions for 
$\gamma = 1.0$, $\alpha = 0.16$.
\item [\ ] {\bf{Figure 6}} The
inertial mass $M_{in}$ and Tolman mass $M_{to}$ for the Melvin branches as
functions of $\gamma$ for $\alpha = 1$. The dashed lines represent the ratio
 $M/\gamma$.
\item [\ ] {\bf{Figure 7}} The approach of the Melvin solution ($\gamma=0$)
by the solutions of the Melvin branch for decreasing values of $\gamma$.
\item [\ ] {\bf{Figure 8}} The quantity $M_{in}(n)/n$ for $n=1$ (solid)
and $n=2$ (dashed) in function of $\alpha $ for $\gamma=1$.
\item [\ ] {\bf{Figure 9}} The profiles of gthe solution on the Kasner branch
for $\gamma=1, \alpha = 0.14$. The function $N$ crosses zero at $x\approx
26$.
\item [\ ] {\bf{Figure 10}} The evolution of the quantity $x_{max}$ on the
Kasner branch in function of $\alpha$ and for $\gamma=1
$.

\end{description}


\newpage
\begin{thebibliography}{7}
\bibitem{verbin} M. Christensen, A.L. Larsen and Y. Verbin, Phys. Rev.
D60 (1999) 12501-2.
\bibitem{niel} H.B. Nielsen and P. Olesen, Nucl. Phys. B61 (1973) 45.
\bibitem{volkov} M.S. Volkov and D.V. Gal'tsov, Phys. Rep. 319 (1999) 1.
\bibitem{bd} Y. Brihaye and M. Desoil, ``Gravitating (bi)-Sphalerons'',
hep-th/0001100.
\bibitem{km} F.R. Klinkhamer and N.S. Manton, Phys. Rev. D30 (1984) 2212.
\bibitem{bm} R. Bartnik and J. Mc Kinnon, Phys. Rev. Lett. 61 (1988) 141.
\bibitem{rebbi} L. Jacobs and C. Rebbi, Phys. Rev. D19 (1979) 4486.
\bibitem{frolov} V. P. Frolov , W. Israel and W. G. Unruh, Phys. Rev. D39 (1989) 1084.
\bibitem{jeannerot} R. Jeannerot, Phys. Rev. Lett. 77 (1996) 3292.
\bibitem{tcha} J. Burzlaff, A. Chakrabarti and D. H. Tchrakian,
J. Phys. A 27 (1994) 1617.
\bibitem{arthur} K. Arthur, Y. Brihaye and D. H. Tchrakian,
Journ. Math. Phys. 39 (1998) 3031.
\end{thebibliography}
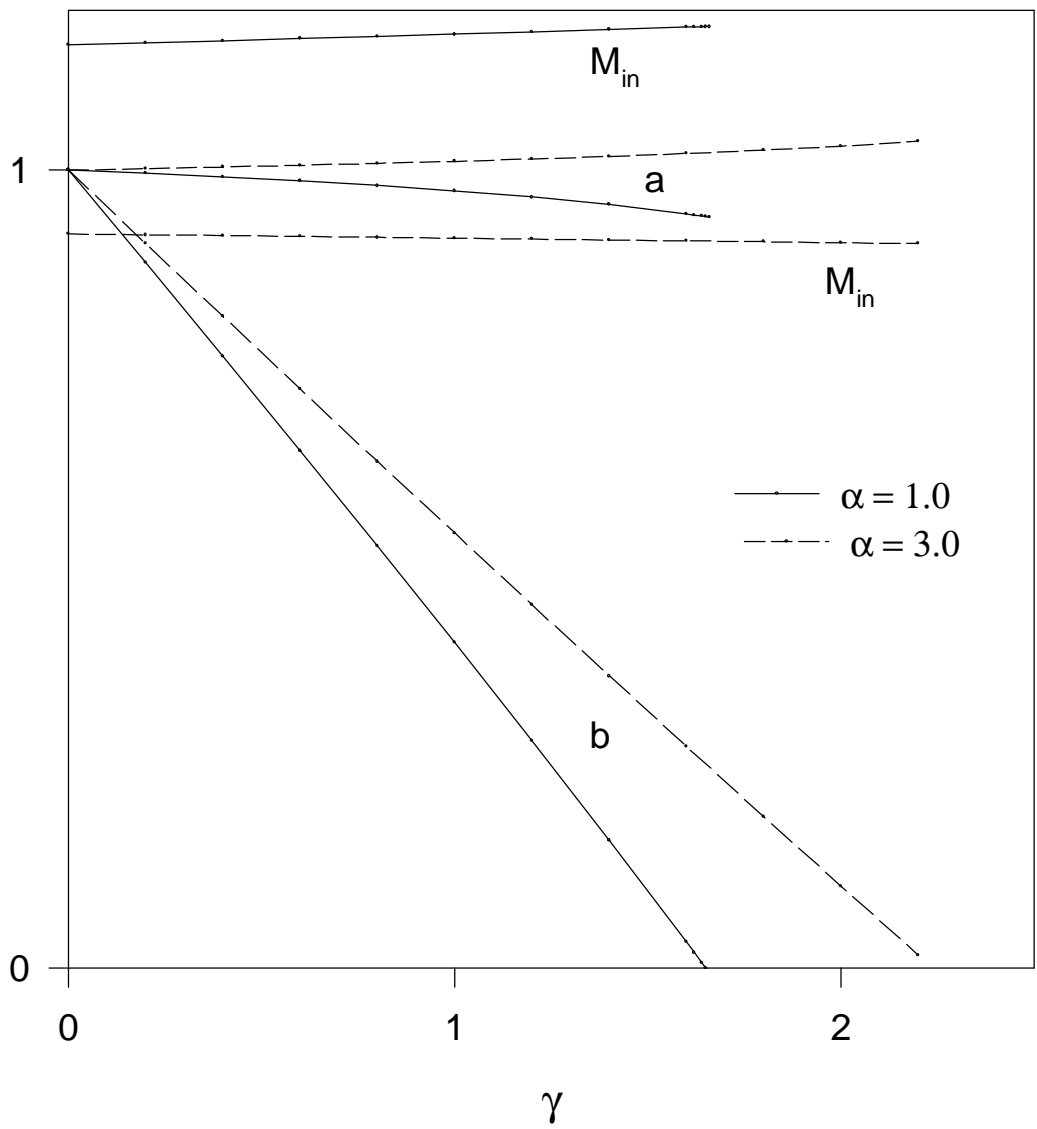
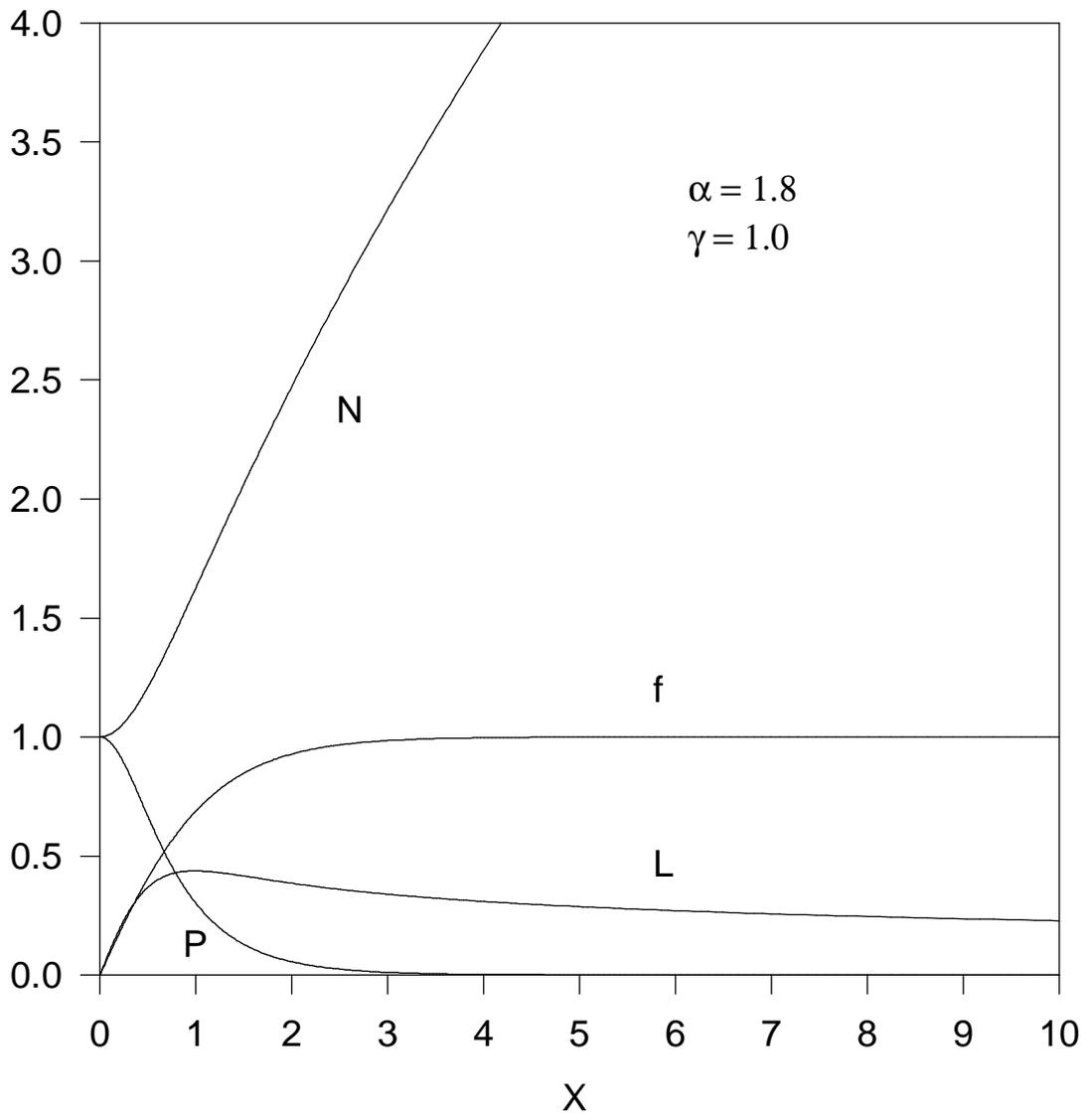
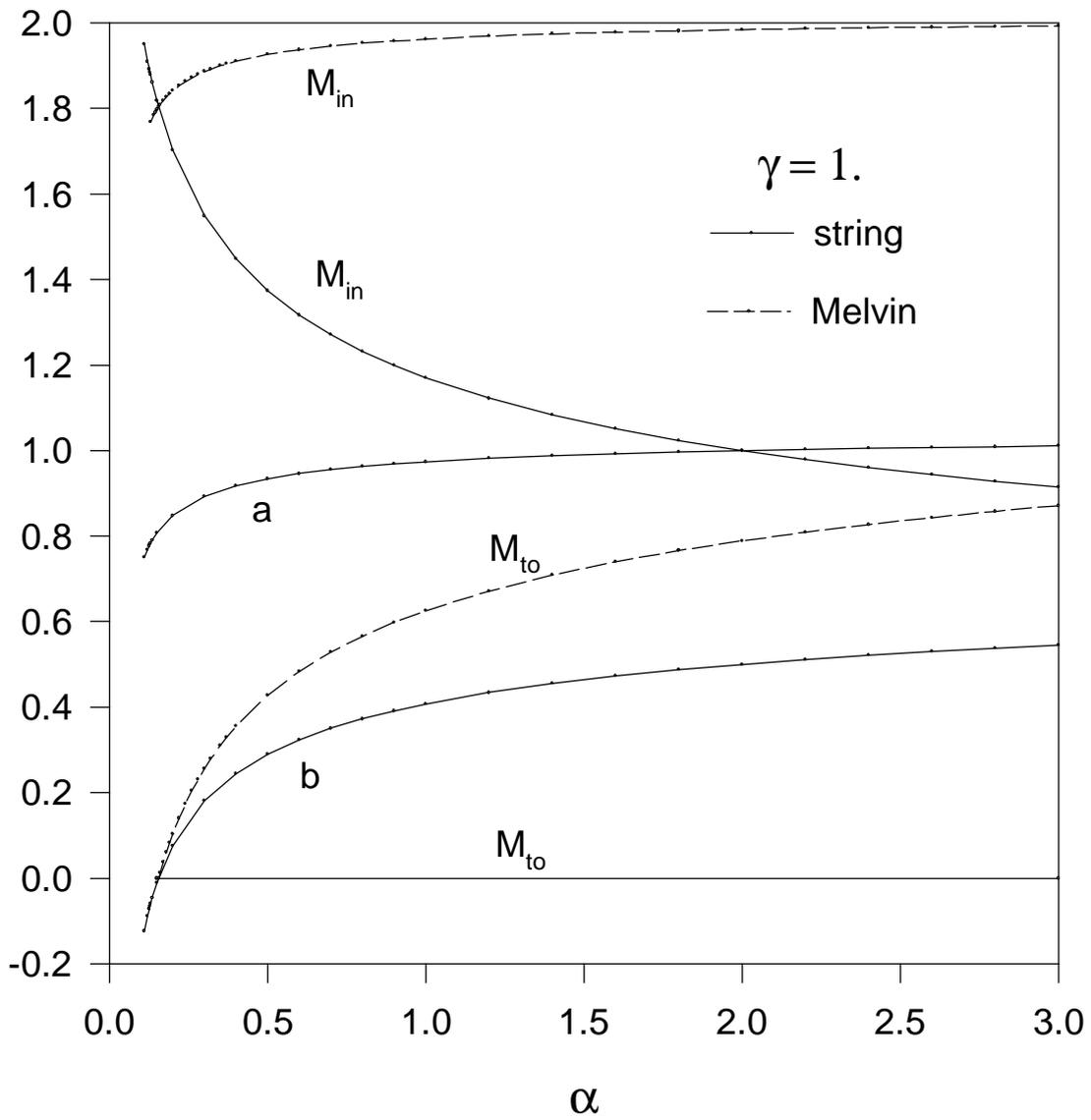
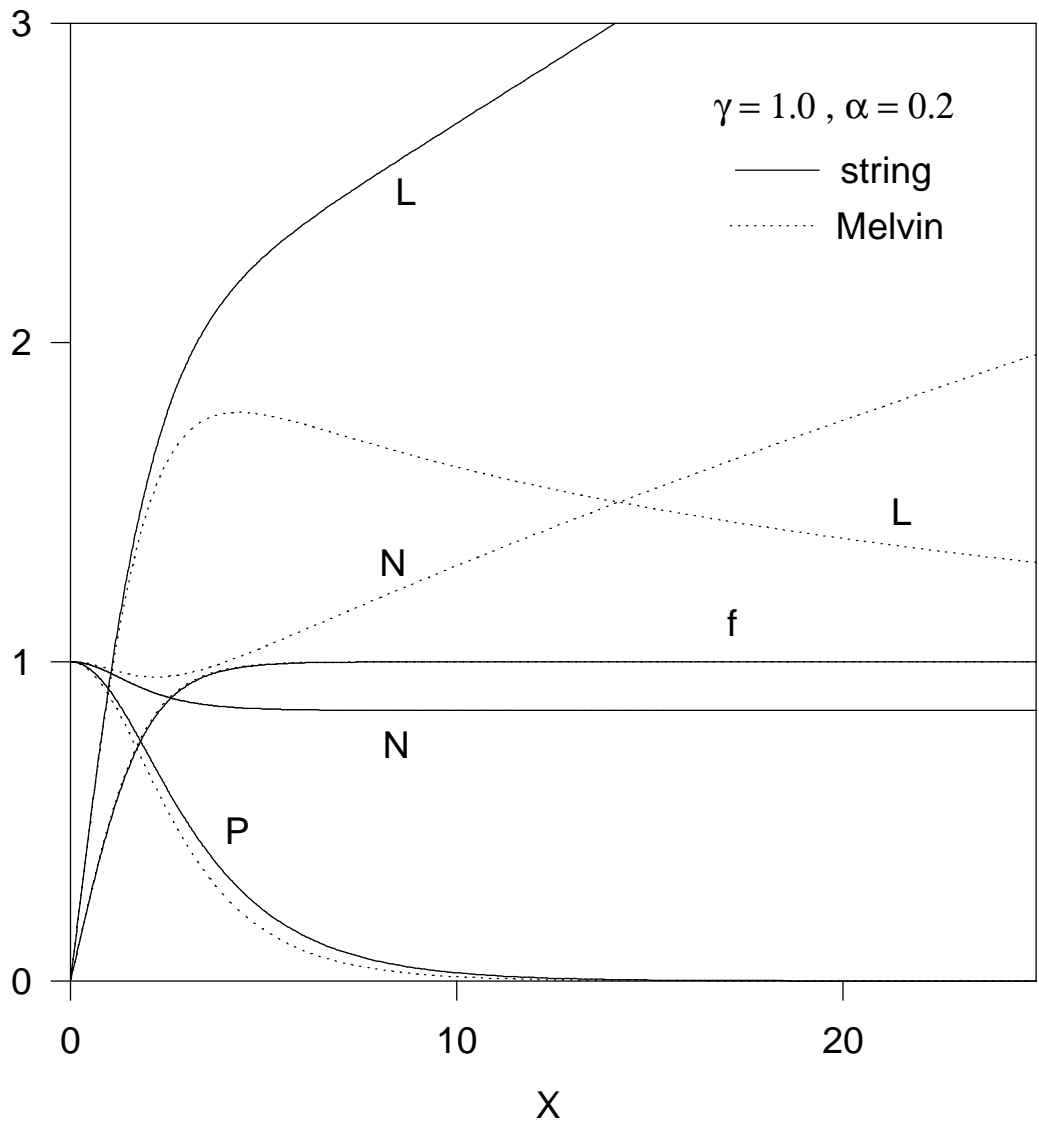
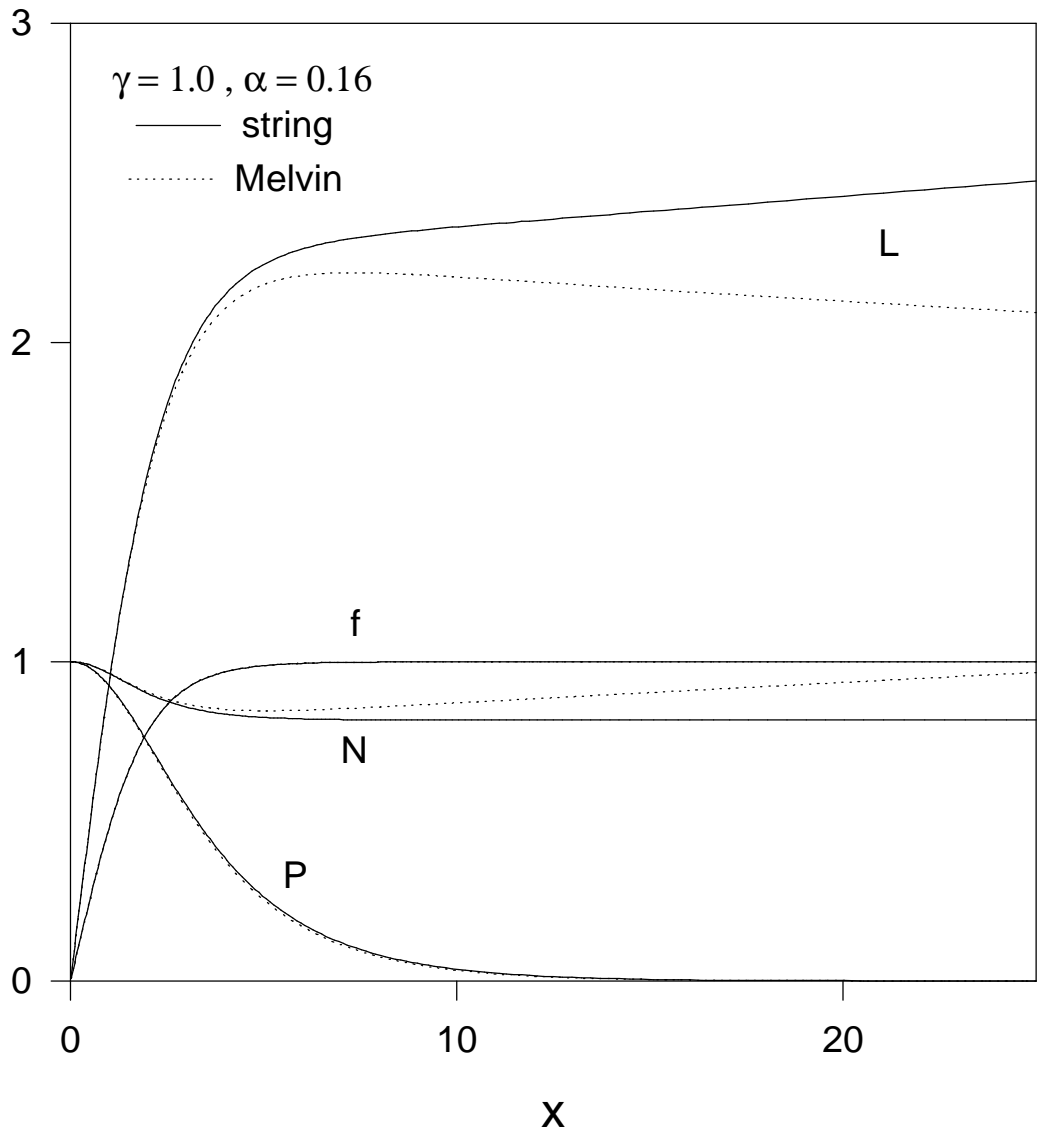
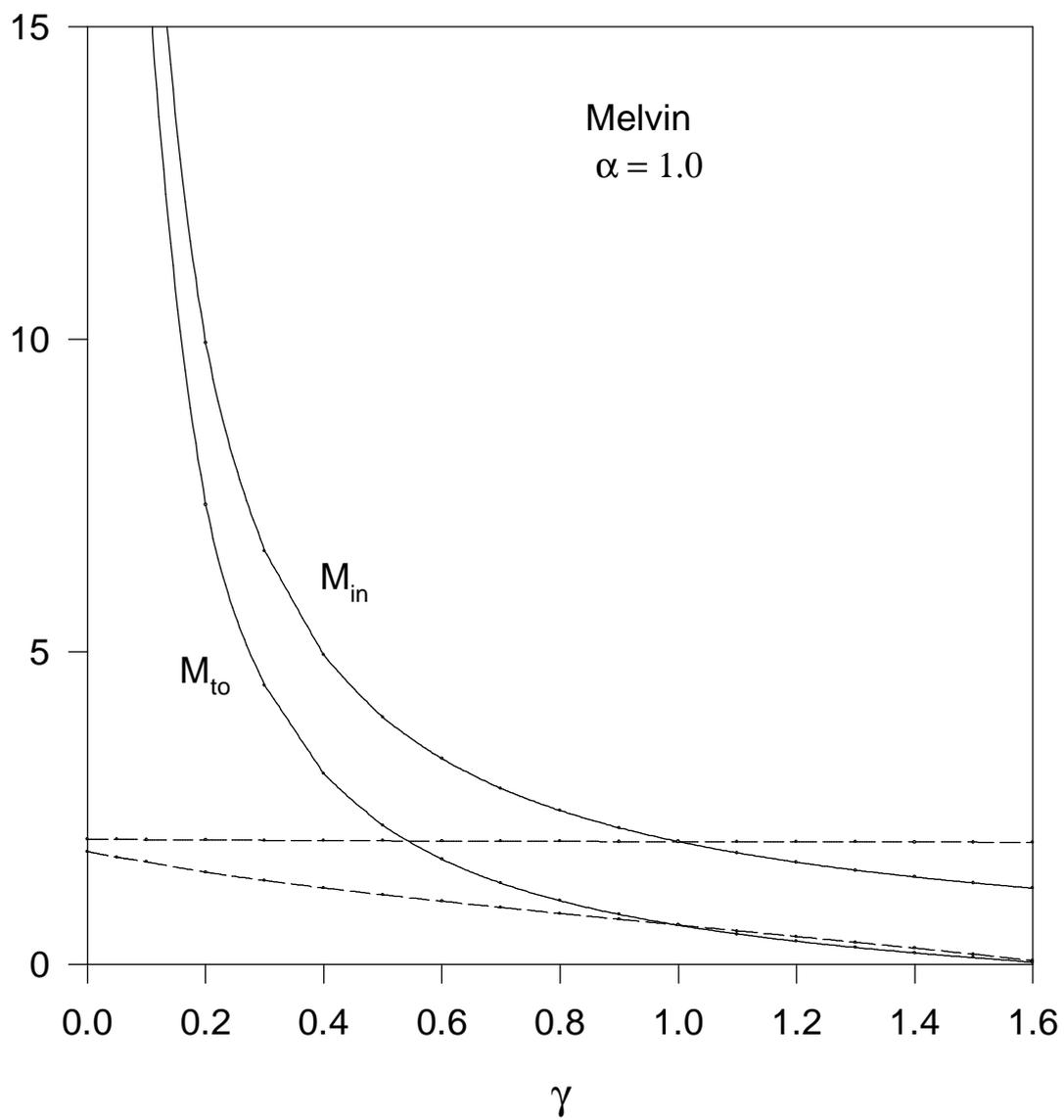
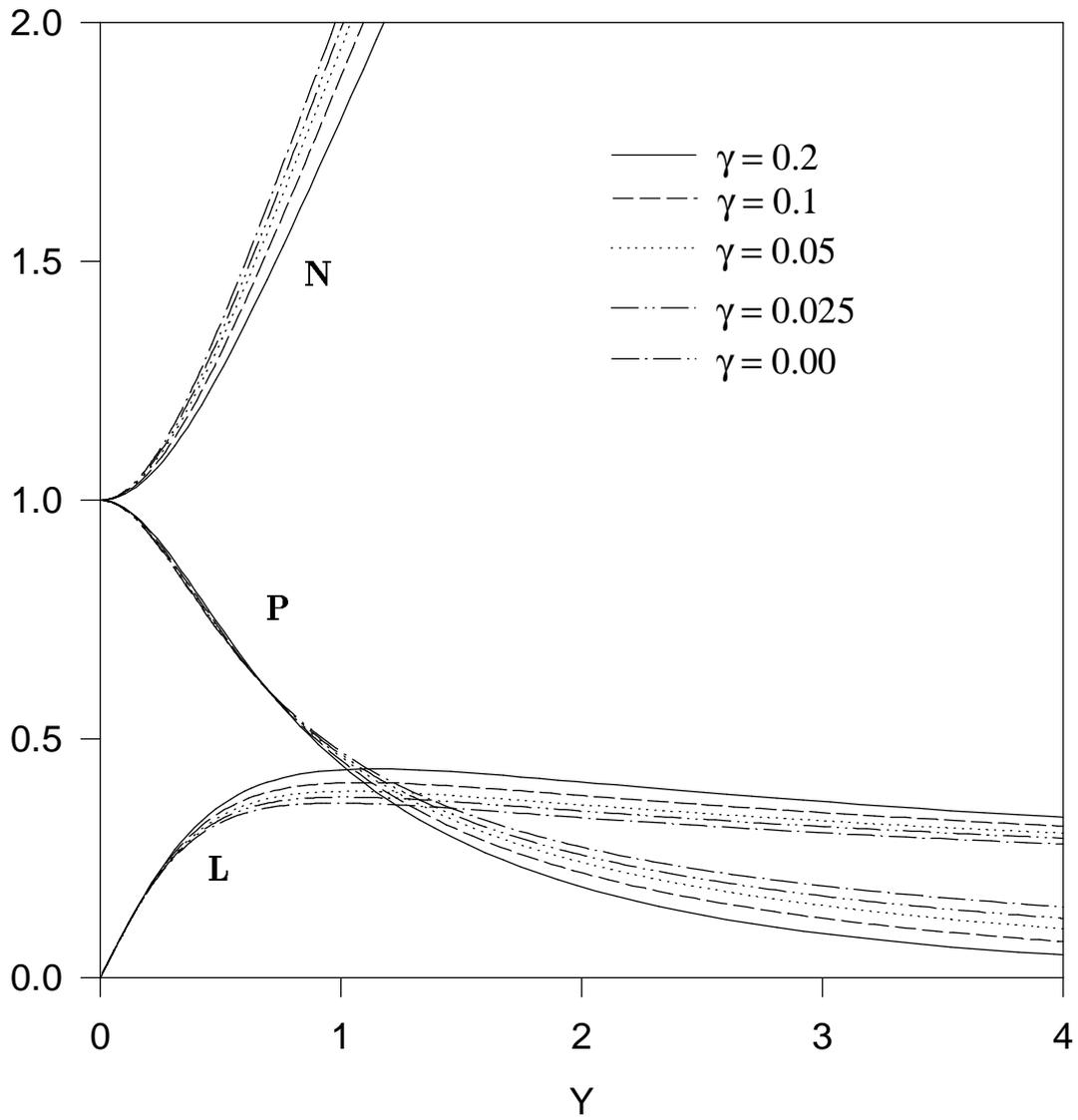
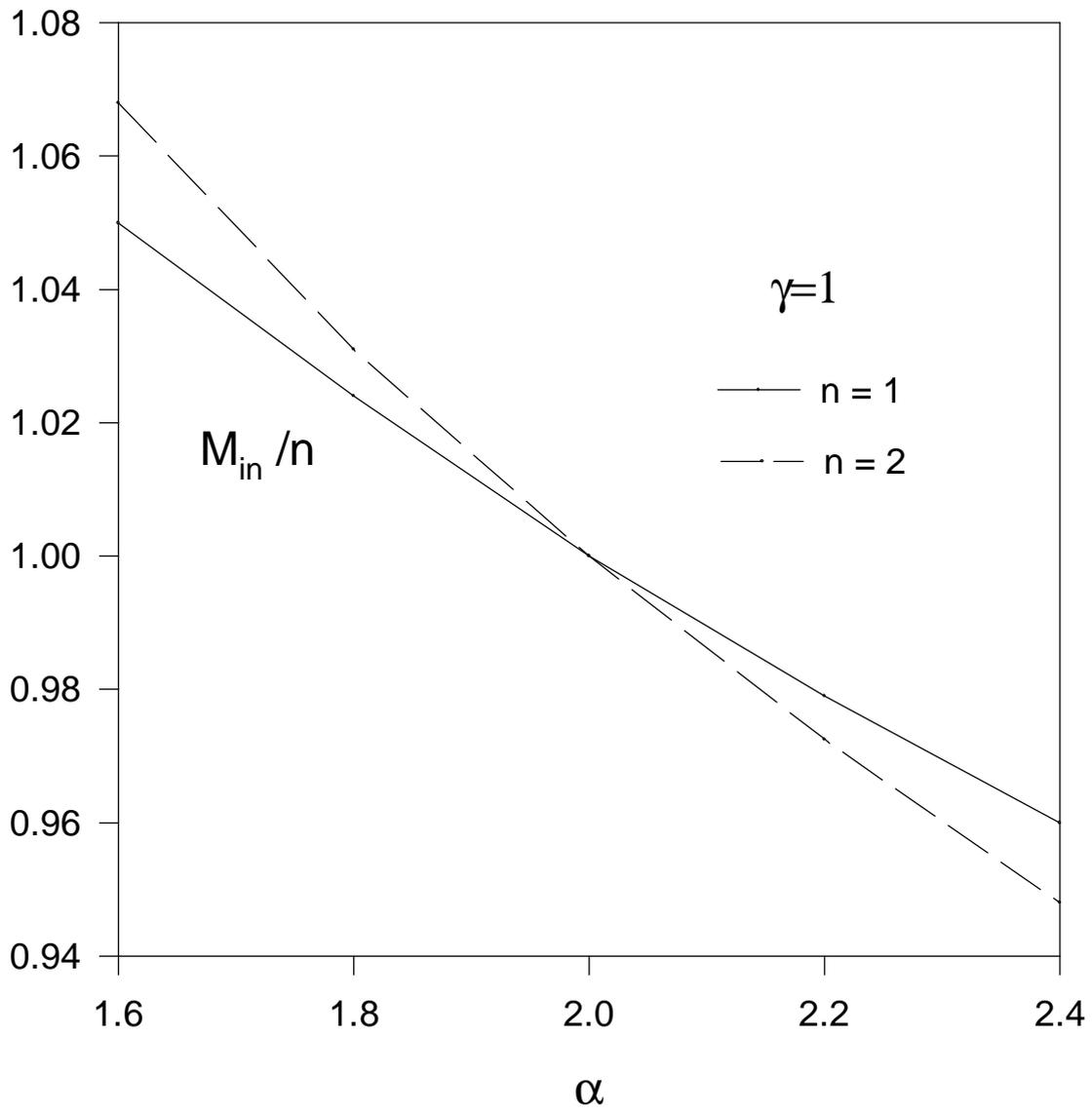
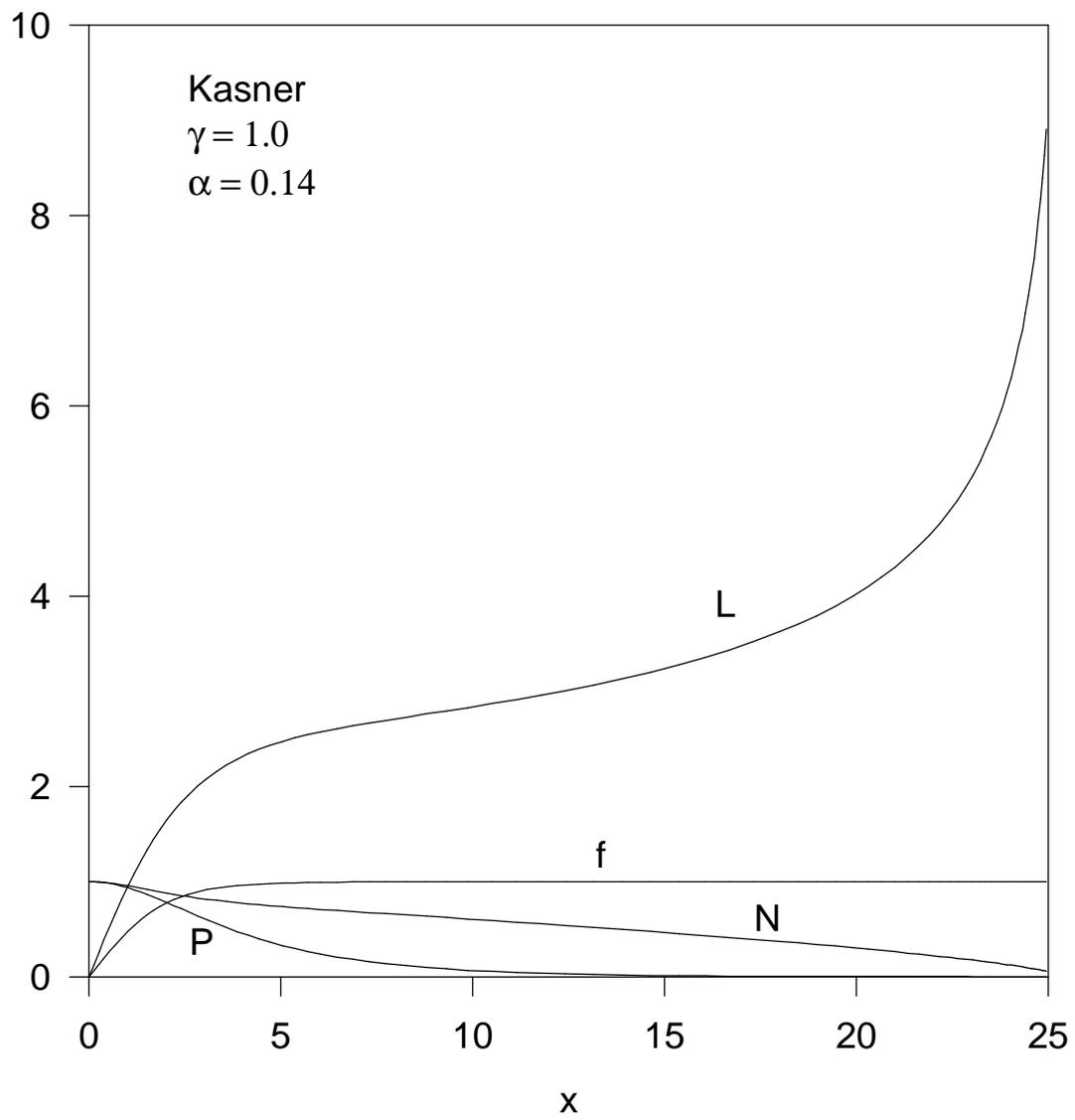
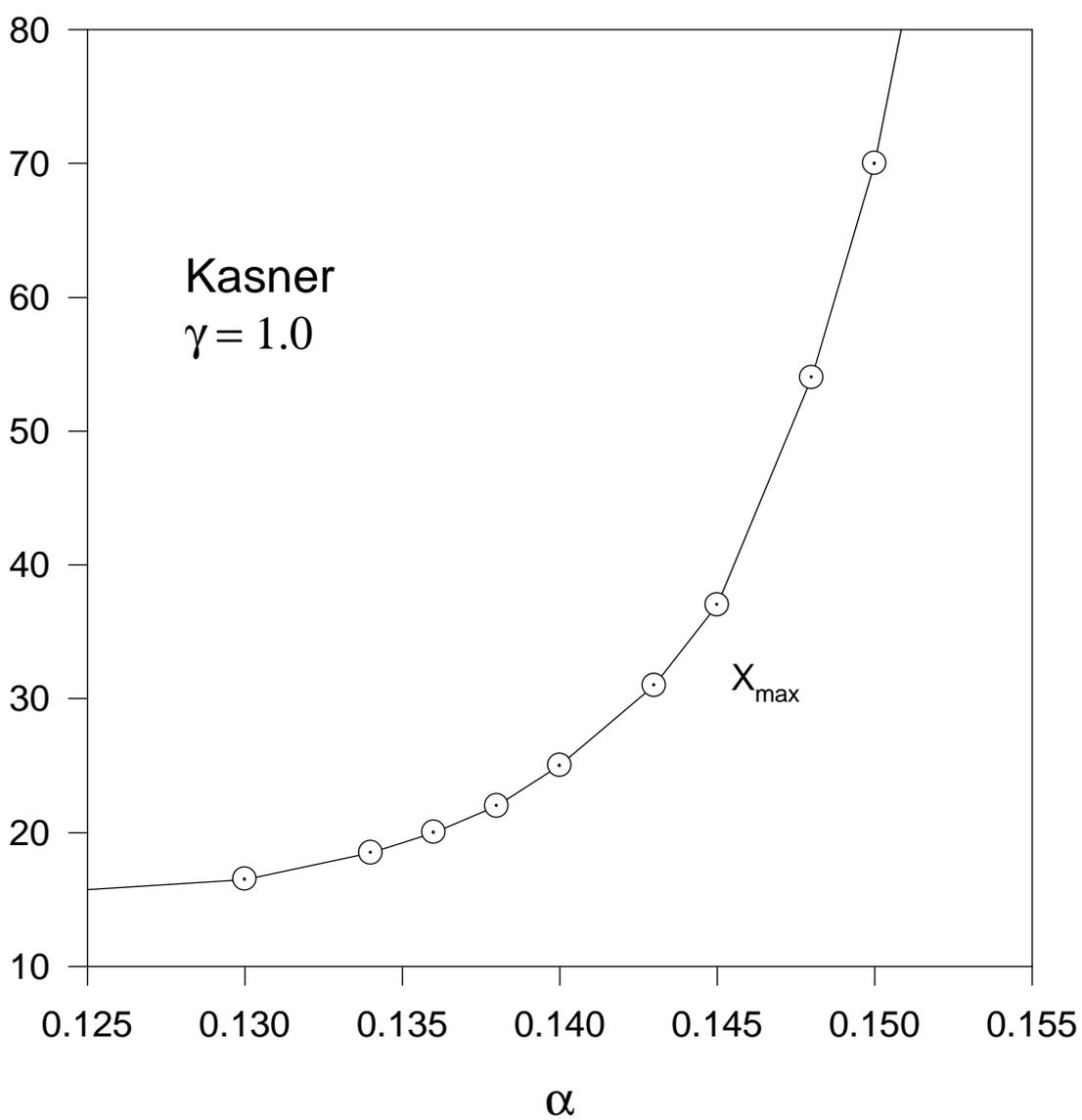
\end{document}